# Social Interactions Clustering MOOC Students: An Exploratory Study

Lei Shi, Alexandra Cristea, Ahmad Alamri
Department of Computer Science, Durham University
Durham, UK
{lei.shi, alexandra.i.cristea,
ahmed.s.alamri}@durham.ac.uk

Armando M. Toda, Wilk Oliveira
Institute of Mathematics and Computer Science
University of Sao Paulo
São Carlos, SP, Brazil
{armando.toda, wilk.oliveira}@usp.br

*Abstract*—An exploratory study on social interactions of MOOC students in FutureLearn was conducted, to answer "*how can we cluster students based on their social interactions?*" Comments were categorized based on how students interacted with them, e.g., how a student's comment received replies from peers. Statistical modelling and machine learning were used to analyze comment categorization, resulting 3 strong and stable clusters.

*Keywords-learning analytics; clustering; social interaction*

## I. INTRODUCTION

E-learning has been capitalizing on motivational theories [1], [2]. Social constructionism [3] views learning as a social process where students construct knowledge for one another and create a small culture of collaboration, sharing resources and thinking. The investigation of the social aspect has shown both benefits and (negative) side effects [4], [5]. This research topic has become more attractive and challenging along the advent of MOOCs (Massive Open Online Courses) due to the effects of massive scale participation [6].

Here, we investigate the social aspect of the FutureLearn MOOCs platform that facilitates social interactions such as discussions and following peers. In FutureLearn, discussions occur when students comment on learning material pages. They can post, reply to and "like" comments. In this study, we focus on commenting behavior and exploring how to cluster students in a semantic way. This is essential as to inform earlier interventions for improvement of student engagement in online courses with social features. Thus, we aim to answer the research questions: *how can we cluster students based on their social interactions?*

## II. RELATED WORK

This section focuses on recent studies analyzing students' social interactions in MOOCs. Sunar et al. [7] explored how social interactions affect learning behavior in FutureLearn, and suggested (1) dropout rates could be reduced by engaging students in social interactions; (2) students interacting with peers were more likely to complete the course. Yang et al. [8] and Shi et al. [9] studied students' behavioral changes along online courses, considering social interactions (comments) as one of the three strong indicators for student clustering. Wang et al. [10] proposed a prediction model regarding learning gains associated with social interactions in MOOCs, concluding that students who presented Active Discourse (students who actively repeated what was in learning material) and Constructive Discourse (students who created content such as explanations and examples) were significant to predict learning gains.

## III. DATASET AND VARIABLE ENGINEERING

The FutureLearn course, "Shakespeare and his World", under investigation was 10-week long. There were 12~17 steps per week (130 in total). A week had an assessment step containing 12 questions, thus 12×10=120 questions in total. The rest 120 steps contained either an article or a video, and students could post, reply to, "like" comments on these steps.

In total, the students posted 44,650 comments: 18,743 (41.98%) were replies to other comments; and the rest, 25,907 (58.02%), were initializing comments, 5,996 (23.14%) of which received at least one reply. On average, a comment received 5 replies (Mean=5.46, SD=51.48). These 44,650 comments were posted by 2,302 "social students".

In this study, we define "social students" as those students who posted at least one comment, but they didn't necessarily post comment(s) every week. On average, a "social student" posted 19.36 comments (Mean=19.36, SD=46.99).

Comments were categorized, based on how the students interacted with them, into *ice-breaking*, *responding* and *solo* (see TABLE I). *Solo* comments were the most (44.59%), while *ice-breaking* the least (13.43%). The majority (76.85%) of the comments did not receive any reply. However, those comments, which did receive replies, received on average 3 (Mean=3.13, SD=6.20). This suggests that a small percentage of students played a dominant role in social interactions.

TABLE I. TYPES OF COMMENTS

| Comment type | Definition | Amount |
|---|---|---|
| **Ice-breaking** | Received at least one reply. | 5,996 (13.43%) |
| **Responding** | Replied to another comment. | 18,743 (41.98%) |
| **Solo** | Not received any reply. | 19,911 (44.59%) |

As shown in TABLE II, the kurtosis value of each type is positive and very large, indicating the distribution has much heavier tails and much sharper peak than normal distributions. Those positive skewness values indicate a skewed non-normal distribution. *Solo* comments have the largest mean with the highest standard deviation (the greatest spread in the data), whilst *ice-breaking* comments have the lowest mean with the lowest standard deviation. On average, a "social student" posted *solo* comments the most; and *solo* comments deviated from the mean the most. Additionally, a Spearman's Rank correlation coefficient test showed weak uphill correlations between each type of comments.

TABLE II.  NUMBER OF DIFFERENT TYPES OF COMMENTS, AS PER STUDENT

|  | *Ice-breaking* | *Responding* | *Solo* |
|---|---|---|---|
| Mean | 4.82 | 9.40 | 17.10 |
| StDev | 7.19 | 15.16 | 47.93 |
| Kurtosis | 25.45 | 12.21 | 67.17 |
| Skewness | 4.08 | 3.21 | 6.94 |

## IV. CLUSTERING, VALIDATION AND DISCUSSIONS

K-means [11] was used to cluster those 2,302 "social students". A big challenge of k-means is to determine the optimal $k$, i.e. the number of clusters. The "elbow method" [12] is a visual method commonly used for this purpose, but the "elbow" cannot always be unambiguously identified. Indeed, we couldn't conclude an optimal $k$ via the "elbow method", so we ran k-means clustering with different $k$s, starting from $k=2$, with further increments of 1. As there were 3 variables in total, we tested $k \in \{n | 2 \leq n \leq 2^3\}$. Before clustering, to make the cluster analysis less sensitive to the scale of the variables, and to be able to compare variables across clusters, we standardized the variables.

When $k=2$ (2 clusters), the convergence was achieved in the 15th iteration. Figure 1 shows the final cluster centers with the number of cases in each cluster. On the standardized scale, all the 3 variables of Cluster 1 were lower than those of Cluster 2, which means these 2,302 "social students" were clustered into a less socially engaged students, i.e. Cluster 1, and a more socially engaged students, i.e. Cluster 2. Besides, Cluster 1 contained much more (2,143; 93.09%) "social students" than Cluster 2 (159; 6.81%). It means the majority of "social students" (in Cluster 1) posted fewer (comparing to "social students" in Cluster 2) comments in each type.

When $k=3$ (3 clusters), the convergence was achieved in the 16th iteration. We can observe, from Figure 2, the final clusters centers, that, similar to the k-means analysis result when $k=2$, the majority (2,082; 90.44%) of those "social students" were allocated in Cluster 3, the least socially engaged cluster, i.e. they posted the fewest comments in each type. Cluster 1 (43; 1.87%) and Cluster 2 (177; 7.69%) represented the more socially engaged clusters. However, comparing Cluster 1 and Cluster 2, the "social students" in the former more likely responded to other students' comments and their comments also received more replies.

When $k = 4$ (4 clusters), the convergence was achieved in the 9th iteration. The final cluster centers (Figure 3) shows, again, the majority (2,065; 89.70%) were allocated in Cluster 3, where comments of each type were posted the least – the students in Cluster III were the least socially engaged. The other 3 clusters represent more socially engaged clusters: Cluster 1 (186; 8.08%) represented those "social students" whose comments received the least replies; Cluster 2 (34; 1.48%) represented those who mainly replied to other's comments instead of initializing a conversation. Cluster 4 (17; 0.74%) represented those who more likely started a conversation and replied to other comments.

When $k = 5$ (5 clusters), the convergence was achieved in the 12th iteration, but one cluster only contained 4 (0.17%) students – too underrepresented comparing to the others, so we decided not to consider 5 as a potential optimal $k$. When $k=6, 7$ and $8$, the convergence could not be achieved within 30 iterations, so we discarded those options. Therefore, the candidate optimal $k$ was $k \in \{n | 2 \leq n \leq 4\}$. However, when $k=2, 3$ and $4$, all clustering results suggested a division between those students who posted fewer comments of each type (less socially engaged) and those students who posted more comments (more socially engaged); while $k \in \{n | 3, 4\}$, the results revealed additional information as to how those students were less/more "socially engaged". Therefore, we discarded $k=2$, and further determined $k$ from $3$ and $4$.

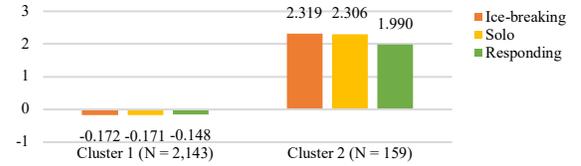

Figure 1. Final cluster centers (when k = 2).

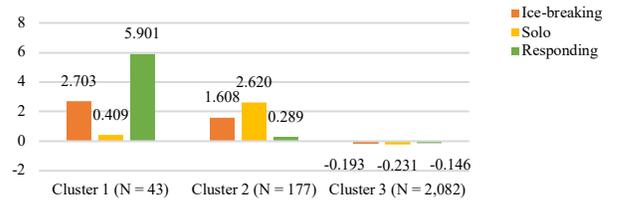

Figure 2. Number of activities – step completions, across weeks.

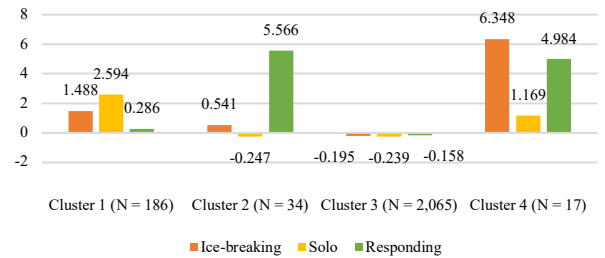

Figure 3. Final cluster centers (when k = 4).

To examine the option of $k = 3$, we conducted a Kruskal-Wallis H test to compare those three clusters in terms of the number of *ice-breaking*, *solo*, and *responding* comments. The result suggests statistically significant differences: (1) *ice-breaking* ($\chi^2 (2) = 398.74, p < .001$), (2) *solo* ($\chi^2 (2) = 434.54, p < .001$), and (3) *responding* ($\chi^2 (2) = 248.75, p < .001$).

We further conducted three Mann-Whitney U tests to compare each pair of clusters, i.e. Cluster 1 versus Cluster 2, Cluster 1 versus Cluster 3, and Cluster 2 versus Cluster 3. The result (TABLE III) suggests that, for each pair, all three variables, i.e. the number of *ice-breaking*, *solo*, and *responding* comments, differ significantly between each other. This indicates that 3 could be the optimal $k$ value.

Similarly, a Kruskal-Wallis H test and six pairwise Mann-Whitney U tests were conducted, under the condition of $k = 4$. The Kruskal-Wallis H test result showed significant differences between four clusters: (1) *ice-*breaking ($\chi^2 (2) = 414.62, p < .001$), (2) *solo* ($\chi^2 (2) = 467.94, p < .001$), and (3)

*responding* ($\chi^2$ (2) = 271.57, $p$ < .001). However, Manny-Whitney U tests' results showed a few pairs didn't differ significantly, i.e. $p \geq 0.05$: (1) the number of *solo* comments did not differ significantly between Cluster 2 and Cluster 3 ($p$ = .97 > .05), and between Cluster 2 and Cluster 4 ($p$ = .06 > .05); (2) the number of *responding* comments did not differ significantly between Cluster 2 and Cluster 4 ($p$ = .06 > .05). Thus, we discarded the option of $k = 4$.

TABLE III. MANN-WHITNEY U TEST RESULT (WHEN K=3)

|  |  | *Ice-breaking* | *Solo* | *Responding* |
|---|---|---|---|---|
| Cluster 1 vs Cluster 2 | U | 3,437.5 | 1,097.0 | 1.5 |
|  | *p* | *0.049* | *<.001* | *<.001* |
| Cluster 1 vs Cluster 3 | U | 15,057.5 | 37,928.0 | 31.5 |
|  | *p* | *<.001* | *0.04* | *<.001* |
| Cluster 2 vs Cluster 3 | U | 33,440.5 | 13,347.0 | 93,352.0 |
|  | *p* | *<.001* | *<.001* | *<.001* |

Thus, we concluded the optimal $k = 3$, and partitioned those 2,302 "social students" into three *strong and stable* clusters. Figure 4 compares the number of *ice-breaking*, *solo* and *responding* comments between these three clusters.

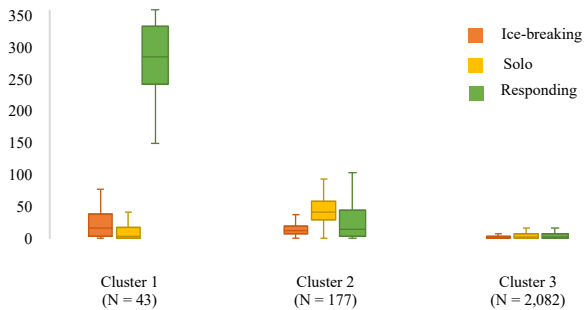

Figure 4. Comparisons of ice-breaking, solo and responding comments.

In sum, we have answered the research question, *how can we cluster students based on their social interactions?* We have found 3 influential variables to determine the clustering, namely, the number of *ice-breaking*, *solo*, and *responding* comments; and, using these three variables, we have found three *strong and stable* clusters to partition these 2,302 "social students", and the students who were allocated to one cluster statistically differ from those who were allocated to the other two clusters. We portray these "social students" as follows:

1. Cluster 1, **Extroverts**: students in this cluster posted much more *ice-breaking* comments to start a conversation; they also posted exceptionally more *responding* comments to continue a conversation; they did not have a lot of *solo* comments as their comments could more likely receive replies.
2. Cluster 2, **Attempters**: students in this cluster did not post as many *ice-breaking* and *responding* comments as extroverts did, but they posted much more comments than the "introverts" did. Although they made effort to start a conversation, a significant number of comments they posted did not receive any reply, i.e. *solo* comments.
3. Cluster 3, **Introverts**: students in this cluster posted the least number of comments in each category. Although they were considered social students by definition in this study, they only occasionally posted comments.

## V. CONCLUSIONS

We have presented an exploratory study on student social interactions in a course delivered on FutureLearn, a MOOCs platform facilitating social learning. We have used statistical modelling and unsupervised machine learning, i.e. k-means, for (1) *variable engineering*, and (2) *clustering and validation*.

In response to the research question, *how can we cluster students based on their social interactions?* we engineered 3 variables, i.e. the number of *ice-breaking*, *solo* and *responding* comments, that contributed to clustering "social students" in the course. Through the clustering analysis, we have found 3 *distinctive* clusters: **Extroverts**, **Attempters** and **Introverts**, which can differentiate students from the perspective of how they socially interact with peers via comments and replies.

This study contributes to the methodology of behavior analysis in MOOC. It shows the potential of social interactions to analyze student behavior, which is essential, as it can be used to inform earlier interventions for engagement especially under the effects of massive scale participation in MOOCs.